\documentclass[12pt,halfline,a4paper]{ouparticle}

\usepackage[utf8]{inputenc} 
\usepackage[T1]{fontenc}    
\usepackage{hyperref}       
\usepackage{url}            
\usepackage{booktabs}       
\usepackage{float}
\usepackage{multirow}
\usepackage{multicol}
\usepackage{nicefrac}       
\usepackage{microtype}      
\usepackage{lipsum}
\usepackage{fancyhdr}       
\usepackage{amsfonts, amssymb, amsmath, amsthm, mathtools, bm} 
\usepackage{graphicx}       
\usepackage[dvipsnames]{xcolor}
\usepackage{dsfont}

\usepackage[round,authoryear]{natbib}

\begin{document}

\title{Tensor-product interactions in Markov-switching models}

\author{
\name{Jan-Ole Koslik}
\address{Bielefeld University, Department of Business Administration and Economics, Bielefeld, 33615, Germany}
\email{jan-ole.koslik@uni-bielefeld.de}
}

\abstract{
Markov-switching models are a powerful tool for modelling time series data that are driven by underlying latent states. 
As such, they are widely used in behavioural ecology, where discrete states serve as proxies for behavioural modes and enable inference on latent behaviour driving e.g.\ observed movement. 
To understand drivers of behavioural changes, it is common to link model parameters to covariates, with nonparametric approaches having gained traction in this context to avoid unrealistic parametric assumptions.
Existing methods are largely limited to univariate smooth functions of covariates 
while real processes are typically complex, requiring 
interaction effects.
We address this gap by incorporating tensor-product interactions into Markov-switching models, enabling flexible modelling of multidimensional effects in a computationally efficient manner. Based on the extended Fellner-Schall method, we develop an 
automatic smoothness selection procedure that is robust and scales well with the number of smooth functions. 
The method builds on a random effects view of the spline coefficients and yields a recursive penalised likelihood procedure.
As special cases, this general framework accommodates bivariate smoothing, function-valued random effects, and space-time interactions.
We demonstrate its practical utility through ecological case studies of an African elephant, common fruitflies, and Arctic muskoxen. The methodology is implemented in the \texttt{LaMa} \texttt{R} package, providing applied ecologists with an accessible and flexible tool to fit 
models with hundreds of parameters and 10-20 (potentially bivariate) smooths. 
}

\date{\today}


\maketitle


\section{Introduction}

In recent years, Markov-switching models --- also referred to as hidden Markov models (HMMs) or latent Markov models --- have gained popularity for analysing time series data characterised by underlying latent states. These models have found applications across various fields, including finance \citep{zhang2019high} and medicine \citep{amoros2019}, but they have gained particular prominence in behavioural ecology, where the discrete states often serve as proxies for distinct behavioural modes \citep{mcclintock2020uncovering}. 
Their popularity in ecology is largely driven by the increasing availability of high-resolution sensor data, allowing researchers to study behavioural patterns based on noisy measurements in a natural environment at an unprecedented scale \citep{nathan2022big}. This capability makes HMMs a valuable tool for understanding how animals interact with their environment and conspecifics and how they respond to changing conditions.

A key advancement in the last decade has been the ability to link state transition probabilities to covariates, providing deeper insights into how hidden processes evolve over time in response to internal and external drivers \citep{patterson2009classifying, patterson2017statistical}. However, modelling these relationships is challenging because the latent states cannot be directly observed, and state inference is only possible \textit{after} fitting a model. Consequently, applied researchers cannot rely on exploratory data analyses to inform model specification.
To address this issue, penalised splines have been introduced as a flexible approach to model complex covariate effects without imposing restrictive parametric assumptions \citep{langrock2015nonparametric, langrock2017markov, langrock2018spline}. Fairly convenient implementation of such models with efficient data-driven smoothness selection has only recently been made available, relying on a random effects view of the spline coefficients and marginal likelihood methods \citep{michelot2022hmmtmb, koslik2024efficient}.

An additional challenge in applied statistical ecology is the presence of individual heterogeneity in multi-animal data sets, motivating the inclusion of random effects to either account for, or directly investigate, individual heterogeneity \citep{gimenez2010individual, hertel2020guide}.
The inclusion of such individual-specific random effects in Markov-switching models has been rather challenging \citep{altman2007mixed, mcclintock2021worth}, but the same recent advances as alluded to above, such as the \texttt{hmmTMB} \texttt{R} package \citep{michelot2022hmmtmb} now facilitate the straightforward incorporation of simple random intercepts using penalised-spline machinery. Additionally, \citet{koslik2024efficient} proposed an alternative method for smoothness selection and variance parameter estimation in an approximate restricted likelihood setting. 
Both of these 
approaches leverage automatic differentiation enabled by the \texttt{R} packages \texttt{TMB} \citep{kristensen2015tmb} and \texttt{RTMB} \citep{kristensen2024rtmb}, respectively, to enable efficient and robust estimation of such semiparametric HMMs.

Despite these major advancements, existing methods remain limited to what we will call \textit{simple} smooths, such as univariate smooths, i.i.d.\ random effects (e.g.\ random intercepts), and isotropic smoothing \citep{wood2003thin}.
Efficient and robust inference procedures for more general smoothing approaches, widely used in modern regression analyses, are still lacking. 
To bridge this gap, we propose an extension of Markov-switching models that incorporates tensor-product interactions of simple smooths. We demonstrate how these interactions facilitate bivariate anisotropic smoothing, function-valued random effects, and space-time interactions, building on concepts from distributional regression \citep{kneib2019modular}. While this extension introduces computational challenges due to more complex penalty structures and high-dimensional parameter spaces, it offers substantial practical benefits because in ecological applications, environmental conditions and behavioural responses can vary across multiple spatial and temporal scales.

To arrive at an efficient and robust estimation scheme, our approach combines three key ingredients: 1) the well-known fast recursive schemes for direct likelihood evaluation in HMMs, 2) automatic differentiation enabled by the novel \texttt{R} package \texttt{RTMB}
, allowing for fast and accurate gradient computation of the HMM likelihood, and 3) efficient automatic smoothness selection via approximate restricted likelihood methods. 
The use of automatic differentiation is particularly crucial, as tensor-product interactions typically lead to high-dimensional parameter spaces where gradient approximations based on finite-differencing are both computationally expensive and numerically unstable.
While the \texttt{RTMB} package supports very general random effects structures, including tensor-product interactions in theory, its efficiency relies on exploiting sparsity in second-derivative calculations. Unfortunately, Markov-switching models lack such sparsity due to temporal dependence in the data, necessitating a tailored smoothness selection procedure.

Specifically, we adopt the extended Fellner-Schall method \citep{fellner1986robust, schall1991estimation} developed by \citet{wood2017fellnerschall}, a generalisation of the method already used by \citet{koslik2024efficient}. 
This method treats the spline coefficients as random effects with an improper multivariate normal distribution, such that smoothness selection can be based on the restricted likelihood of the smoothing parameters, with all other model parameters integrated out using the Laplace approximation.
Our implementation is fully modular, allowing seamless integration of tensor-product interactions into custom Markov-switching models specified using simple \texttt{R} code. Compared to naive grid-search approaches, the iterative nature of this method offers substantial computational advantages by scaling efficiently with the number of smoothness parameters and enabling a stable transition from highly penalised and hence very stable, to more flexible models, thereby mitigating numerical issues such as convergence to local optima.

The paper is structured as follows. Section \ref{sec:model formulation} introduces the basic HMM model formulation and motivates the inclusion of tensor-product interactions. Section \ref{subsec:tp_interactions} outlines the construction of tensor-product interactions, focussing on the special cases mentioned above. 
Section \ref{sec:smoothness} presents the extended Fellner-Schall method for smoothness selection. Lastly, Section \ref{sec:case studies} illustrates our approach with three case studies, demonstrating bivariate smoothing of time-of-day and day-of-year-dependent transition probabilities, function-valued random effects, and space-time interactions.


\section{Basic model formulation and inference tools}

\label{sec:model formulation}

A hidden Markov model (HMM), also called Markov-switching model, comprises two stochastic processes: an observed process $\{X_t\}_{t = 1, \dotsc, T}$ and a latent, unobserved state process $\{S_t\}_{t = 1, \dotsc, T}$, the latter taking values in the finite set $\{ 1,\ldots, N\}$. The model is mainly characterised by two dependence assumptions. First, at any given time $t$, conditional on the value of $S_t$, the observation $X_t$ is independent of all previous and future values of both the observed process and the state process, which is formally known as the \textit{conditional independence assumption}. Thus, the conditional distribution of $X_t$ is fully specified by the density (or probability mass function) $p_i(x_t) = p(x_t \mid S_t = i)$ for $i = 1, \dotsc, N$. 
Second, the state process $\{S_t\}$ is assumed to be a first-order Markov chain, hence fully characterised by its initial state distribution $\bm{\delta}^{(1)} = \bigl(\Pr(S_1 = 1), \dotsc, \Pr(S_1 = N)\bigr)$ and the possibly time-varying transition probability matrix (t.p.m.)
$$\boldsymbol{\Gamma}^{(t)}=(\gamma_{ij}^{(t)}), \; \text{ with } \; \gamma_{ij}^{(t)} = \Pr(S_{t}=j \mid S_{t-1}=i), \quad t = 2, \dotsc, T.$$
Thus, the model is fully specified by the parameters governing the state-dependent distributions and the parameters characterising the state process. We denote the collection of all model parameters by the vector $\bm{\theta}$, with appropriate link functions applied to each element such that $\bm{\theta}$ is unconstrained in $\mathbb{R}^d$. 

For such an HMM, parameter estimation can be conducted efficiently by using a recursive scheme called the \textit{forward algorithm}. This algorithm exploits the Markov property to sum over all possible latent state sequences efficiently, with a computational complexity that scales linearly in the number of observations $T$ \citep{zucchini2016hidden, mews2024build}. Written in closed form, the recursive scheme for likelihood evaluation is
\begin{equation}
    \label{eqn:likelihood}
    \mathcal{L}(\bm{\theta}) = \bm{\delta}^{(1)} \bm{P}(x_1) \bm{\Gamma}^{(2)} \bm{P}(x_2) \bm{\Gamma}^{(3)} \dotsc \bm{\Gamma}^{(T)} \bm{P}(x_T) \bm{1},
\end{equation}
where $\bm{P}(x_t) = \text{diag}\bigl(p_1(x_t), \dotsc, p_N(x_t)\bigr)$ is a diagonal matrix containing the state-dependent densities evaluated at observation $x_t$ and $\bm{1}$ is a column vector of ones.
In practice, \eqref{eqn:likelihood} suffers from numerical underflow or overflow even for moderate $T$ but a slight modification of the algorithm permits stable recursive calculation of the log-likelihood $\ell(\bm{\theta}) = \log \mathcal{L}(\bm{\theta})$ (for more details, see \citealp{zucchini2016hidden}). 
In principle, $\ell(\bm{\theta})$ can then be optimised using any standard Newton-Raphson-type numerical optimisation routine.
Within this flexible inference framework, any model parameter --- influencing $\bm{\delta}^{(1)}$, $\bm{\Gamma}^{(t)}$ or $\bm{P}(x_t)$ --- can be linked to covariates via linear predictors and appropriate link functions (see \citealp{michelot2022hmmtmb}, \citealp{koslik2024efficient}, for more details). 

Due to the flexibility of the inference framework, incorporating the broad class of penalised splines into Markov-switching models is relatively straightforward \citep{langrock2015nonparametric, langrock2017markov, feldmann2023flexible} and particularly attractive due to their mathematical simplicity \citep{langrock2018spline}. 
There are several options to include penalised splines in such models. For example, instead of choosing a particular parametric family of state-dependent distributions, univariate densities be expressed as linear combinations of fixed basis functions:
$$
p_i(x_t) = \sum_{k=1}^K \alpha_k^{(i)} B_k(x_t),
$$
where the $\alpha_k^{(i)}$ are non-negative weights summing to one, the $B_k$ are normalised B-spline basis functions that integrate to one and a suitable penalty is placed on the $\alpha_k^{(i)}$ to prevent overfitting \citep{eilers1996flexible, langrock2015nonparametric}.
Furthermore, generalised additive models (GAMs) can be incorporated either in the state process or state-dependent process of HMMs. 
For the former, transition probabilities can be expressed as functions of linear predictors using the inverse multinomial logistic link function
$$\gamma_{ij}^{(t)} = \frac{\exp(\eta_{ij}^{(t)})}{\sum_l \exp(\eta_{ik}^{(t)})},$$
or a Markov-switching GAM can be obtained by letting the state-dependent distributions depend on time-varying parameters, for example,
$$p_i^{(t)}(x_t) = p(x_t; \mu_i^{(t)}, \phi_i), \quad h(\mu_i^{(t)}) = \eta_i^{(t)},$$
for some parametric density $p$, dispersion parameter $\phi_i$, and a suitable invertible link function $h$ \citep{langrock2017markov}. 
In both cases, the linear predictors take the form
$$
\eta^{(t)} = \beta_0 + f_1(z_{t1}) + \dotsc + f_Q(z_{tq}),
$$
where the $f_q$ are smooth functions of the covariates $z_{tq}$ which have a representation in terms of a linear combination of fixed basis functions, and we omitted state indices for notational simplicity. 
Practically, this means that the model is defined by a set of design matrices for the linear predictors and a set of penalty matrices that penalise the smoothness of the functions $f_q$. A more detailed explanation of model estimation by penalised likelihood is given in Section \ref{sec:smoothness}.
For a more exhaustive summary of popular model formulations incorporating penalised splines and case studies involving each of these examples, see \citet{koslik2024efficient}.

From an applied perspective, including such nonparametric effects is particularly desirable in Markov-switching models as the dependence on the unobserved state sequence considerably complicates model formulation. As the hidden states can only be inferred after a model is fitted to data, exploratory data analysis, for example, to gain intuition on the relationship of the transition probabilities on covariates, is not possible. Furthermore, selecting the number of hidden states is notoriously difficult \citep{pohle2017selecting} because in applications one is typically faced with misspecification due to the immense complexity of the real processes being modelled. 
When guided by information criteria, practitioners tend to compensate for such misspecification by adding more states, but this often makes the model unnecessarily complicated --- while the actual reason for the misspecification might actually lie in unrealistic parametric assumptions for the covariate effects.
While there has been some progress recently in reliable selection of the number of states under misspecification \citep{hung2013hidden, de2024estimation, dupont2024improved}, including flexible nonparametric relationships might still be a superior option to potentially uncover relationships that otherwise would have been overlooked. 
While the ability to include smooth univariate functions of covariates is highly beneficial, there are many practical situations where relationships may vary with additional covariates or change over time --- scenarios that cannot be adequately captured using only univariate smooth functions, motivating the use of tensor-product interactions.

\section{Tensor-product interactions and associated penalties}
\label{subsec:tp_interactions}


Generalising the examples from the previous section, in an HMM any parameter can, in principle, be linked to covariates via a suitable linear predictor and link function. Hence, for the subsequent section, consider a generic parameter $\nu$ which could be a state-dependent parameter or which could (partly) specify the transition matrix. 
Consider now that $\nu$ should depend on covariates $z_{t1}$ and $z_{t2}$ via
$$
h(\nu_t) = \eta_t = \beta_0 + f(z_{t1}, z_{t2}),
$$
where $h$ is a suitable bijective link function and $f$ is an arbitrary smooth function, with the meaning of \textit{smooth} to be discussed. 
In this section, we omit the covariates' time index $t$ for ease of notation.
Following \citet{kneib2019modular}, we express $f$ in terms of fixed basis functions $B_k(z_1, z_2)$ as
$$
f(z_1, z_2) = \sum_{k=1}^K \beta_k B_k(z_1, z_2) = \bm{\beta}^\intercal \bm{B}(z_1, z_2),
$$
where $\bm{B}(z_1, z_2) = \bigl(B_1(z_1, z_2), \dotsc, B_K(z_1, z_2)\bigr)^\intercal$ is the vector of basis function evaluations and $\bm{\beta} = (\beta_1, \dotsc, \beta_K)^\intercal$ is a vector of associated coefficients. 
To enforce $f$ to be smooth, it is common to penalise the basis coefficients by adding a penalty of the form
$$
-\lambda \bm{\beta}^\intercal \bm{S} \bm{\beta}
$$
to the log-likelihood function, where $\bm{S}$ is a fixed penalty matrix that depends on the basis used and our operationalisation of \textit{smoothness}. 
For the two-dimensional basis of interest here, we could, for example, choose a thin-plate regression spline basis, which would yield the fixed basis functions $B_k$ and associated penalty matrix $\bm{S}$. While incorporating a two-dimensional function, such a predictor still yields a ``simple'' smooth with a single smoothing parameter and could be estimated using the tools developed by \citet{michelot2022hmmtmb} or \citet{koslik2024efficient}.

With this model formulation, one can, in principle, represent interactions of covariates by choosing appropriate basis functions. However, we cannot freely choose a basis for both marginal smooths separately, and also there is only a single smoothness parameter $\lambda$ associated with smoothness across both dimensions. The latter is an unrealistic assumption in many real scenarios. For example, if $z_{1}$ is the time of day and $z_{2}$ is the julian day, then there is no mechanistic reason why the smoothness of diel variation should be similar to that of the variation measured over the scale of one year. Hence, it is useful to construct the basis functions and penalty matrices separately for both marginal effects and construct the interaction from there.

Accordingly, assume the marginal smooth effects of both covariates can be represented as
$$
f_1(z_1) = \sum_{k_1 = 1}^{K_1} \beta_{1 k_1} B_{1 k_1}(z_1), \qquad f_2(z_2) = \sum_{k_2 = 1}^{K_2} \beta_{2 k_2} B_{2 k_2}(z_2),
$$
where $B_{1k_1}$ and $B_{2k_2}$ are determined by some arbitrary basis, with associated penalty matrices $\bm{S}_1$ and $\bm{S}_2$. Similar to the parametric case, the interaction of the two marginal smooths can be obtained by considering all pairwise products of basis functions, leading to the representation of $f$ as
\begin{equation}
\label{eqn:tensorproduct}
f(z_1, z_2) = \sum_{k_1=1}^{K_1} \sum_{k_2=1}^{K_2} \beta_{k_1 k_2} B_{1 k_1}(z_1) B_{1 k_2}(z_2).
\end{equation} 
The above specification can be best thought of on a 2-dimensional landscape, where the basis functions $B_{1 k_1}(z_1) B_{1 k_2}(z_2)$ represent hills whose height is controlled by the associated coefficient $\beta_{k_1 k_2}$. Penalisation now needs to be applied across two dimensions, which can be achieved by defining the penalty matrix
\begin{equation}
    \label{eqn:tp_penalty_matrix}
    \bm{S}_\lambda = \lambda_1 (\bm{S}_1 \otimes \bm{I}_{K_2}) + \lambda_2 (\bm{I}_{K_1} \otimes \bm{S}_2) 
\end{equation}
where $\otimes$ denotes the Kronecker product and $\bm{I}_K$ is the identity matrix of dimension $K$. The first term in the summation corresponds to applying the smoothness penalty to each row and the second term to each column of the tensor product coefficients. Critically, using this construction, smoothness is controlled separately for the two dimensions by $\lambda_1$ and $\lambda_2$. For further details, see \citet{wood2017generalized}. Importantly, having two smoothness parameters changes the structure of the penalty from a quadratic form multiplied by a scalar to a more complicated form where the penalty matrix itself depends on two smoothing parameters, making established smoothing parameter selection methods \citep{michelot2022hmmtmb, koslik2024efficient} inapplicable.

Conceptually, it is straightforward to consider higher-dimensional tensor-product interactions, but here we restrict ourselves to the special case of two marginal smooths for simplicity and practical feasibility. For a thorough discussion of the general case, see \citet{wood2017generalized}.

\subsubsection*{ANOVA decomposition of interactions}

In many cases, it is beneficial to decompose a tensor-product interaction into its main effects and a pure interaction term, resulting in a model of the form
$$
f(z_1, z_2) = \beta_0 + f_1(z_1) + f_2(z_2) + f_{1,2}(z_1, z_2).
$$
Such a model can be preferable as it simplifies the interpretation of the estimated effects. Structurally, the function space spanned by this model is identical to that of a tensor product with absorbed main effects, but there are subtle differences. Crucially, such a decomposition introduces additional smoothing parameters for the two main effects. This can indeed be advantageous in cases where components of $f_{1,2}$ should be shrunken towards the main effect rather than to zero --- as is the case for function-valued random effects. 

For a model containing the above predictor to be identifiable, the smooth function $f_{1,2}$ must be constrained to exclude the main effects from its span. \citet{wood2017generalized} shows how imposing such constraints can be fairly straightforward: for the marginal smooths to be identifiable, the constant function needs to be removed from their span. This can achieved by imposing \textit{sum-to-zero} constraints by subtracting the column means from the respective design matrices, i.e.\ centering the columns of the design matrix, which ensures that $f_1$ and $f_2$ are orthogonal to the intercept term. Once the constant function is removed, it follows that the interaction of $f_1$ and $f_2$ cannot include main effects anymore, as this would correspond to the product of the constant function and one of the $f_i$.

\citet{wood2013straightforward, wood2017generalized, kneib2019modular} discuss alternative constructions and constraints in much more detail, but the above specification is sufficient for our purposes. Furthermore, for practical applications, often it is best to use the design and penalty matrices provided by the \texttt{mgcv} \texttt{R} package \citep{wood2015package}, with the convenience that all constraints have already been absorbed into the provided model matrices and need not be worried about.

\subsection*{Important special cases}

\textit{Bivariate tensor-product splines.}
The most straightforward tensor-product interaction arises from combining two univariate penalised splines. For instance, each marginal smooth could be chosen as a cubic regression spline with an associated penalty matrix. The tensor-product representation can then be constructed as outlined above. 
Depending on the application, the interaction can be represented either as a tensor product that includes the main effects or using the ANOVA decomposition, with the latter often providing better interpretability.

\textit{Function-valued random effects.}
Sometimes, data might be grouped by a factor variable $z_1$ with levels  $1, \dotsc, K$, such as when analysing tracks of different individuals.
If there is substantial heterogeneity in responses to a second variable $z_2$, it may be necessary to include a smooth effect of $z_2$ at each level of $z_1$. However, often there is reason to suspect the groups to be somewhat similar and share a common mean effect.

A natural modelling approach is then to consider the interaction of a group-specific random intercept and a smooth function $f(z_{2})$. The random intercept term can be expressed as a simple smooth by letting
$$
B_{1k}(z_1) = \mathds{1}(z_1 = k),
$$
with associated penalty matrix $\bm{S}_1 = \bm{I}_{K}$. Constructing the tensor-product interaction of this term with a smooth function of $z_2$ and applying the ANOVA decomposition yields a \textit{function-valued random effect}. Crucially, this formulation allows for the estimation of separate penalty strength (or equivalently variance) parameters for the smoothness along 
$z_2$ and the random-effect dimension. 
The latter effectively shrinks the interaction term towards zero, reducing each group-specific effect to the main effect if there is little evidence for individual variation in the data. Denoting $f_{1,2}(z_1, z_2)$ by $f_{z_1}(z_2)$ to emphasise the random-effect character, the resulting additive predictor takes the form
$$
\beta_0 + \beta_{k} + f(z_2) + f_{k}(z_2), \quad k = 1, \dotsc, K,
$$
where $\beta_{k}$ is a group-specific random-intercept and $f_{k}(z_2)$ a group-specific function-valued random effect.

\textit{Space-time interactions.}
Indeed, in the characterisation of the tensor-product interaction in  \eqref{eqn:tensorproduct}, $z_{1}$ or $z_{2}$ need not necessarily be univariate.
An important special case arises when $z_1$ consists of two-dimensional spatial coordinates.
In this scenario, a two-dimensional marginal smooth can be defined for $z_1$, for example using the previously-mentioned thin plate regression splines, which are isotropic and governed by a single smoothing parameter associated with the penalty. For spatial coordinate data, isotropy is often a reasonable assumption, as we typically do not expect systematic differences in smoothness between the $x$- and $y$-coordinates of GPS data.
The second variable $z_{2}$ might then represent some kind of temporal variable like the time of the observation or the time of day, whose influence can, for example, be modelled as a cubic regression spline. Forming the tensor-product interaction of these two marginal smooths then yields a \textit{space-time interaction}. Expressing $z_{1} = (x, y)$, such a model then comprises additive predictors of the form
$$
\beta_0 + f_{xy}(x, y) + f_t(t) + f_{xyt}(x, y, t),
$$
where again $f_{xy}(x, y)$ and $f_t(t)$ are simple smooths and $f_{xyt}(x, y, t)$ requires anisotropic smoothing. Again, we might be tempted to choose the ANOVA decomposition approach here, as it facilitates straightforward comparisons to simpler models excluding the temporal effect.


\section{Smoothness selection via the extended Fellner-Schall method}

\label{sec:smoothness}

As discussed in the previous section, a tensor-product smooth can be represented either as a single smooth with two penalty matrices (hence also two penalty parameters) or as a combination of two ``simple'' smooths along with a pure interaction term. In general, a Markov-switching model may include several simple smooths as well as several tensor-product interactions --- potentially one for each state or each off-diagonal entry of the t.p.m. 
For notational simplicity, it is thus helpful to represent the zoo of distinct penalties on various subvectors of the coefficient vector $\bm{\theta}$ by a single penalty on the entire coefficient vector. Consequently, we define the full-model penalty matrix
$$
\bm{S}_{\lambda} = \sum_{j = 1}^L \lambda_j \mathcal{S}_j,
$$
where $L$ is the number of penalty matrices and each $\mathcal{S}_j$ is non-zero only for the indices in $\bm{\theta}$ that are to be penalised by this matrix. Hence, this formulation includes the tensor-product interaction case that is characterised by \textit{overlapping} penalties. As a simple example, consider
$$
\mathcal{S}_1 = \begin{pmatrix}
    \bm{0} & \bm{0} & \bm{0}\\
    \bm{0} & \bm{S}_1 & \bm{0}\\
    \bm{0} & \bm{0} & \bm{0}\\
\end{pmatrix}, \quad
\mathcal{S}_2 = \begin{pmatrix}
    \bm{0} & \bm{0} & \bm{0}\\
    \bm{0} & \bm{0} & \bm{0}\\
    \bm{0} & \bm{0} & \bm{S}_2\\
\end{pmatrix}, \quad
\mathcal{S}_3 = \begin{pmatrix}
    \bm{0} & \bm{0} & \bm{0}\\
    \bm{0} & \bm{0} & \bm{0}\\
    \bm{0} & \bm{0} & \bm{S}_3\\
\end{pmatrix},
$$
which would amount to a model with fixed effects for the first indices of $\bm{\theta}$ corresponding to the upper-left zero block in $\bm{S}_{\lambda}$, then a univariate (or isotropic) smooth with one smoothing parameter and penalty matrix $\lambda_1 \bm{S}_1$, and lastly, a tensor-product interaction with penalty matrix $\lambda_2 \bm{S}_2 + \lambda_3 \bm{S}_3$, where $\bm{S}_2$ and $\bm{S}_3$ are given by the two matrices resulting from the Kronecker products in \eqref{eqn:tp_penalty_matrix}. 

For any given penalty strength parameter (vector) $\bm{\lambda}$, the model can be estimated by optimising the penalised log-likelihood
\begin{equation}
\label{eqn:penalised_llk}
    \ell_p(\bm{\theta}; \bm{\lambda}) = \ell(\bm{\theta}) - \bm{\theta}^\intercal \bm{S}_{\lambda} \bm{\theta} / 2
\end{equation}
numerically, using standard off-the-shelf numerical optimisers. 
However, for efficient and robust computation in such high-dimensional settings, it indeed becomes necessary to use automatic differentiation tools, but this is described in more detail below.

The more challenging task lies in selecting an appropriate smoothing parameter in a data-driven way. To do so, we adopt a random effects view, where the penalty above imposes an improper Gaussian prior on $\bm{\theta}$. Specifally, $\bm{\theta} \sim \mathcal{N}(\bm{0}, \bm{S}_{\lambda}^-)$, where $\bm{S}_{\lambda}^-$ denotes the Moore-Penrose inverse of $\bm{S}_{\lambda}$. Estimation can then proceed by restricted likelihood methods, i.e., integrating out $\bm{\theta}$ --- which contains fixed and random effects --- from the joint likelihood. In a slight abuse of notation, $\Hat{\bm{\lambda}}$ should then be chosen to maximise
\begin{equation}
    \label{eqn:marginal_likelihood}
    \mathcal{L}_r(\bm{\lambda}) = p_{\lambda}(\bm{x}) = \int_{\mathbb{R}^d} p(\bm{x} \mid \bm{\theta}) \: p_{\lambda}(\bm{\theta}) \; d\bm{\theta},
\end{equation}
where $p(x \mid \bm{\theta}) = \mathcal{L}(\bm{\theta})$ and $p_{\lambda}(\bm{\theta})$ is the (improper) Gaussian density associated with the distribution of $\bm{\theta}$. This expression can be interpreted as the average likelihood score achieved upon drawing $\bm{\theta}$ from its prior distribution, considered as a function of $\bm{\lambda}$.
As the high-dimensional integral in \eqref{eqn:marginal_likelihood} is numerically intractable, we perform a \textit{Laplace approximation} --- based on a second-order Taylor expansion of the joint log-likelihood (the logarithm of the integrand in \eqref{eqn:marginal_likelihood}) around its mode $\Hat{\bm{\theta}}_\lambda = {\arg\max}_{\bm{\theta}} \; \ell_p(\bm{\theta}; \bm{\lambda})$, which is obtained by numerically maximising \eqref{eqn:penalised_llk} \citep{wood2016smoothing}. This approximation leads to the restricted log-likelihood
\begin{equation}
\label{eqn:restricted_llk}
    \ell_r(\bm{\lambda}) = \ell(\Hat{\bm{\theta}}_{\lambda}) - 
    \Hat{\bm{\theta}}_{\lambda}^\intercal \bm{S}_{\lambda} \Hat{\bm{\theta}}_{\lambda} / 2 + 
    \log \vert \bm{S}_{\lambda} \vert_+ / 2 - 
    \log \vert \bm{H}_{\lambda} + \bm{S}_{\lambda} \vert /2 + \text{const.},
\end{equation}
where $\bm{H}_{\lambda} = - \partial^2 \ell / \partial \bm{\theta} \partial \bm{\theta}^\intercal$ and $\vert \bm{A} \vert_+$ denotes the product of all non-zero eigenvalues of the matrix $\bm{A}$.

In principle, one could now try to optimise \eqref{eqn:restricted_llk} directly using a quasi-Newton optimiser, but this either requires higher-order derivatives of $\ell(\bm{\theta})$ 
or finite differencing of \eqref{eqn:restricted_llk}, where for each new trial value of $\bm{\lambda}$, the inner optimisation has to be re-run. The \texttt{R} packages \texttt{TMB} and \texttt{RTMB} take the first approach via automatic differentiation, but evaluating $\partial \bm{H}_\lambda / \partial \lambda$ is costly if sparsity cannot be exploited. 
Unfortunately, for HMMs, due to the global temporal dependence introduced by the forward algorithm, $\bm{H}_{\lambda}$ will generally be dense, making automatic differentiation prohibitively slow in the high-dimensional tensor-product setting of interest.
Clearly, the computational cost of the second option is even worse, requiring at least $L$ inner optimisations per outer iteration, where $L$ is the number of smoothing parameters. 

Hence, a further approximation is needed that allows for a good tradeoff between estimation accuracy and computational efficiency. A good candidate is given by the so-called extended Fellner-Schall method \citep{fellner1986robust, schall1991estimation, wood2017fellnerschall}, which proceeds as follows.
Partially differentiating \eqref{eqn:restricted_llk} w.r.t. $\lambda_j$ 
yields
\begin{equation}
\label{eqn:deriv_restricted_llk}
    \frac{\partial \ell_r}{\partial \lambda_j} =
    - \Hat{\bm{\theta}}_{\lambda}^\intercal \mathcal{S}_{j} \Hat{\bm{\theta}}_{\lambda} / 2 +
    \text{tr}(\bm{S}_{\lambda}^- \mathcal{S}_j) / 2 - 
    \text{tr}(\bm{J}_\lambda^{-1} \mathcal{S}_j) /2 -
    \text{tr}(\bm{J}_\lambda^{-1} \partial \bm{H}_{\lambda} / \partial \lambda_j )/2,
\end{equation}
where $\bm{J}_\lambda = \bm{H}_{\lambda} + \bm{S}_{\lambda}$. The term directly involving the log-likelihood vanishes, since $\partial \ell_p / \partial \bm{\theta} \vert_{\Hat{\bm{\theta}}_{\lambda}} = \bm{0}$ by definition of $\Hat{\bm{\theta}}_{\lambda}$. The most problematic part of the above equation is the last term, as it involves the explicit dependence of the Hessian of the negative log-likelihood on the smoothing parameter. However, \citet{breslow1993approximate, gu1992cross, wood2017fellnerschall} neglect the dependence of $\bm{H}_\lambda$ on $\bm{\lambda}$ justified by it vanishing asymptotically.
Having made this approximation, \eqref{eqn:deriv_restricted_llk} (without its last term) can now be used to construct the estimating equation
\begin{equation}
    \label{eqn:lambda_update}
    \lambda_j^* = \lambda_j \frac{\text{tr}(\bm{S}_{\lambda}^- \mathcal{S}_j) - \text{tr}(\bm{J}_{\lambda}^{-1} \mathcal{S}_j)}{\Hat{\bm{\theta}}_{\lambda}^\intercal \mathcal{S}_{j} \Hat{\bm{\theta}}_{\lambda}}.
\end{equation}
When \eqref{eqn:deriv_restricted_llk} is positive, the fraction in \eqref{eqn:lambda_update} is larger than one, thus $\lambda_j$ will increase. When \eqref{eqn:deriv_restricted_llk} is negative, $\lambda_j$ will decrease and once \eqref{eqn:deriv_restricted_llk} is zero, $\lambda_j$ no longer changes. 
Note that if the block in $\bm{S}_{\lambda}$, corresponding to the non-zero elements in $\mathcal{S}_j$, only contains a single penalty matrix --- as it does for simple smooths --- $\text{tr}(\bm{S}_{\lambda}^- \mathcal{S}_j)$ simplifies to $\text{rank}(\bm{S}_j) / \lambda_j$ reducing \eqref{eqn:lambda_update} to the updating equation already presented by \citet{koslik2024efficient}.
The approximation made above is critical as when employing \eqref{eqn:lambda_update} to estimate $\bm{\lambda}$, only the first two derivatives of the log-likelihood are needed, which need be computed anyway to find $\Hat{\bm{\theta}}_\lambda$, while the evaluation of prohibitively costly higher-order derivatives can be avoided.



Comparing the procedure to more naive grid-search approaches, it becomes evident that the iterative nature is extremely valuable. Even if the optimal penalty strength was known, fitting an HMM with the corresponding penalty by numerical optimisation could easily result in convergence to a local optimum. However, if the model fit is initialised with a fairly large penalty strength, the initial inner optimisation is very stable. Subsequent inner optimisations can then be initialised with the penalised estimate $\Hat{\bm{\theta}}_\lambda$ from the previous iteration and the penalisation is only gradually reduced (as determined by \eqref{eqn:lambda_update}). Hence, each inner penalised fit alters the previous estimate $\Hat{\bm{\theta}}_\lambda$ only slightly, typically leading to fast inner convergence. Ultimately, this results in a very stable and efficient procedure, which is immensely valuable in the high-dimensional optimisation setting arising from tensor-product interactions.

\subsection*{Practical implementation}
\label{subsec:practical}

For practical implementation, we provide the two functions \texttt{qreml()} and \texttt{penalty2()} in the \texttt{R} package \texttt{LaMa} \citep{koslikLaMa2024}. 
These build on the \texttt{R} package \texttt{RTMB}
, an \texttt{R} interface to the \texttt{TMB} package, to allow for automatic differentiation in the penalised likelihood estimation --- while not using the package's Laplace approximation functionalities because smoothness selection is based on the approximate gradient of the restricted log-likelihood via \eqref{eqn:lambda_update}.

To use \texttt{qreml()}, the user merely needs to implement a penalised negative log-likelihood function that is compatible with \texttt{RTMB} and use the \texttt{penalty2()} function to compute all quadratic form penalties. The likelihood function can then be passed to \texttt{qreml()}, and after specifying which parameters are spline coefficients (or random effects), the extended Fellner-Schall method is used to find the optimal penalty strength parameter. The outer optimisation is terminated once the maximum component of \eqref{eqn:deriv_restricted_llk} falls below a threshold of $10^{-4}$ in absolute value. 
The inner optimisation is performed by \texttt{optim()} \citep{rstats} with the BFGS method 
because it showed the overall best performance regarding both speed and stability. For each $\bm{\lambda}$ update, the (approximate) Hessian is obtained from \texttt{optim()} by setting \texttt{hessian = TRUE}, which in turn applies finite differencing to the gradient provided by \texttt{RTMB} to approximate the Hessian. Note that \texttt{RTMB} does indeed provide the option to evaluate the Hessian via automatic differentiation, but we found this option to be considerably slower than finite differencing the gradient.

In Theorem 1, \citet{wood2017fellnerschall} state that to guarantee a positive $\lambda_j^*$, $\bm{H}_{\lambda} = \bm{J}_\lambda + \bm{S}_\lambda$ neads to be positive definite. Hence, we include a check whether the observed Hessian is positive definite, and if not, replace it with its nearest positve definite matrix. 
Another important practical consideration is that applying \eqref{eqn:lambda_update} is not guaranteed to increase $\ell_r$, hence it is important to control the step size to make the optimisation more stable and reliable. Therefore, we include an exponential smoothing parameter $\alpha \in (0,1)$, applying the simple rule $\bm{\lambda}^{(k)} = (1-\alpha) \bm{\lambda}^{(k)*} + \alpha \bm{\lambda}^{(k-1)}$, where $\bm{\lambda}^{(k)*}$ is the proposal based on \eqref{eqn:lambda_update}. 
Slowing the outer optimisation to some extent is also benefitial for a second reason: typically, each penalised inner optimisation is complicated and high-dimensional, with a non-negligible potential to converge to a local optimum. Thus, it is advantageous to initialise with a large penalty strength to obtain a stable initial fit and then only \textit{slowly} decrease the penalty strength, making the iterative penalised model fits substantially more stable. By default, $\alpha$ is set to 0.3, which we found to produce reliable results.

\textit{Smoothness parameter uncertainty quantification.} 
Especially for functional random effects, quantifying estimation uncertainty in the smoothness parameters --- or typically the random-effect variance (their inverse) --- is desirable. Conceptually this is straightforward as maximum likelihood theory states that approximately 
$\Hat{\bm{\lambda}} \sim \mathcal{N}(\bm{0}, \mathcal{H}_r^{-1})$, where $\mathcal{H}_r = \partial^2 \ell_r/ \partial \bm{\lambda} \partial \bm{\lambda}^\intercal \vert_{\Hat{\bm{\lambda}}}$, but again, obtaining the second-derivative of the restricted likelihood is computationally expensive in practice as it involves the derivative of the Hessian (in $\bm{\theta}$) w.r.t.\ $\bm{\lambda}$, again making this option infeasible for the dense high-dimensional setting under consideration. Hence, we propose an approximation based on finite differencing of \eqref{eqn:deriv_restricted_llk} (without the term depending on $\partial \bm{H}_\lambda / \partial \lambda_j$), which is implemented in the function \texttt{sdreport\_outer()}. This, in turn, needs $J$ additional penalised model fits if $J$ is the dimension of $\bm{\lambda}$. If uncertainty quantification for $\sigma_j^2 = 1 / \lambda_j$ is desired, the function can also apply the delta method.

\textit{Parameter mapping.} The simple nature of the updating equation presented in \eqref{eqn:lambda_update} allows for the following convenient feature. Consider the case that several smooths with indices in $I \subset \{1, \dotsc, J\}$ are assigned the same smoothing parameter $\lambda_I$. In this case, to obtain an updating equation for $\lambda_I$ it suffices to sum all terms in the numerator and denominator with indices contained in $I$. This is exploited in the \texttt{qreml()} function by allowing users to pass an optional \texttt{map} argument. This argument is a list, with the component named after the penalty parameter containing a factor vector that forces selected smoothness parameters to be the same. For example, if a model has four smoothing parameters, \texttt{factor(c(1,1,2,2))} leads to the estimation of only two unique parameters, shared by smooths one and two as well as three and four, respectively. Additionally, components set to \texttt{NA} are fixed to their initial value and excluded from estimation.


\section{Case studies}

\label{sec:case studies}

The following three case studies demonstrate smoothness selection for tensor-product interactions in Markov-switching models based on the extended Fellner-Schall method outlined in the previous section. Code for reproducing the case studies can be found at \url{https://github.com/janoleko/tp_interactions}.

\subsection{African elephant --- year-round diurnal variation}

To demonstrate the feasibility of including bivariate smooth functions of covariates in Markov-switching models, we analyse the movement track of an African elephant (\textit{Loxodonta}) from the Ivory Coast. The dataset consists of 12,170 longitude and latitude GPS recordings, collected every two hours between September 2018 and November 2021, and is freely available in the Movebank repository 2736765655 \citep{movebank2024}.
As is common practice when analysing movement data derived from GPS locations, the GPS positions were converted into \textit{step lengths} (km) and \textit{turning angles} (radians) \citep{langrock2012flexible}.

To these data, we fitted 2-state hidden Markov models (HMMs), assuming state-dependent gamma distributions for the step lengths and von Mises distributions for the turning angles, with conditional independence between the two observed variables given the underlying state. While a 2-state model may be an overly simplistic representation of the elephant's behaviour, the relatively coarse temporal resolution does not allow for more detailed inference regarding its behavioural process. An initial fit with a homogeneous Markov chain as the state-process model suggested that the first state is characterised by slow, undirected movement, whereas the second state involved larger steps and more directed movement. Accordingly, we interpret the two states as ``encamped'' and ``exploratory'' behaviour (cf. \citealp{morales2004extracting}).

To investigate the elephant's behavioural diel variation and the seasonal variation therein, we fit a model involving a tensor-product interaction of the two cyclic variables, relating them to both off-diagonal entries of the transition probability matrix. Formally,
$$
\text{logit}(\gamma_{ij}^{(t)}) = \beta_0^{(ij)} + f_{\text{tday}}^{(ij)}(\text{tday}_t) + f_{\text{julian}}^{(ij)}(\text{julian}_t) + f_{\text{tday}, \text{julian}}^{(ij)}(\text{tday}_t, \text{julian}_t), \quad i \neq j ,
$$
where $\text{tday}_t$ and $\text{julian}_t$ denote the time of day and the day of year, respectively.

As a baseline comparison, we also fit models comprising only one of the two effects and one model comprising a purely additive effect.
For the largest model, involving the tensor-product interaction, we use the ANOVA decomposition explained in Section \ref{subsec:tp_interactions} to separate the smooth function into main effects and a pure interaction term. 
The model matrices for the interaction model were obtained using \texttt{mgcv} with the formula
\begin{center}
\texttt{s(tday, bs = "cc", k = 12) + s(julian, bs = "cc", k = 12) +\\
ti(tday, julian, bs = "cc", k = 12).}
\end{center}

\begin{table}
    \centering
    \begin{tabular}{lccc}
    \toprule
        transition probability formula & log-likelihood & $\Delta$ AIC & $\Delta$ BIC\\
    \midrule
        $\sim$ \texttt{1} & -28757.38 & 2990.70 & 2777.50\\
        $\sim$ \texttt{s(tday)} & -27355.09 & 217.41 & 120.16\\
        $\sim$ \texttt{s(julian)} & -28687.61  & 2864.75 & 2701.76\\
        $\sim$ \texttt{s(tday) + s(julian)} & -27248.33 & 23.75 & \textbf{0.00}\\
        $\sim$ \texttt{s(tday) + s(julian) + s(tday, julian)} & -27221.71 & \textbf{0.00} & 85.47\\
    \bottomrule
    \end{tabular}
    \caption{Log-likelihood values and information criteria of the five models fitted to the elephant data.}
    \label{tab:elephant_IC}
\end{table}

In total, the model with the tensor-product interaction comprises 248 coefficients (the cyclic bases results in $10 + 10 + 10^2 = 120$ coefficients for each linear predictor) and eight smoothness parameters that need to be estimated --- four for each off-diagonal entry of the t.p.m. These smoothness parameters were initialised with $10^4$ for the main effects and $10^5$ for the interaction. Model estimation then took about seven minutes on an Apple M2 chip with 16 GB of memory. The extended Fellner-Schall method required a total of 22 penalised fits for convergence of the outer optimisation, i.e.\ until the largest component of the approximate outer gradient fell below $10^{-4}$. Estimation of the other candidate models was substantially faster, taking under one minute each, so we do not report the exact times here.

Table \ref{tab:elephant_IC} reports AIC and BIC values for the five candidate models. Notably, these are \textit{conditional} AIC and BIC values, based on the unpenalised log-likelihood of the model at the final penalised optimium and the corresponding effective number of parameters \citep{gray1992flexible}, not \textit{marginal} AIC and BIC, which could also be obtained using the random effects representation.
AIC favours the most complex model while BIC prefers the simpler additive model.
In general, selecting HMMs based solely on information criteria is not recommended \citep{pohle2017selecting}; instead, the choice should be guided by the specific research question. Therefore, we examine the results of the most complex model, which includes the interaction term, while acknowledging that the simpler additive model could also be a viable choice.

\begin{figure}
    \centering
    \includegraphics[width=1\linewidth]{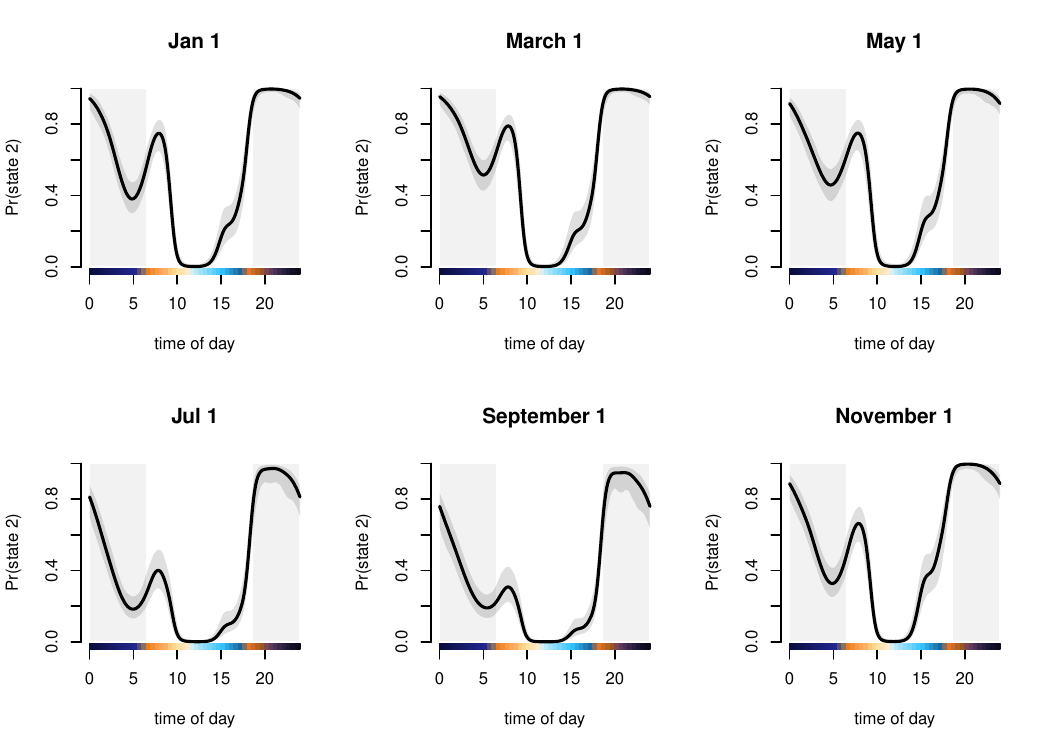}
    \caption{Probability of the elephant being in the exploratory state at different times of the day for six different months. Pointwise 95\% confidence intervals were obtained by simulating from the approximate posterior distribution of $\Hat{\bm{\theta}}$.}
    \label{fig:elephant_stationary}
\end{figure}

From this model, we can now compute the periodically stationary state distribution \citep{koslik2023inference} for the daily cycle, holding the day of the year constant. Doing this at various days of the year, we can investigate the change in diel variation over the year. 
Based on this state distribution, Figure \ref{fig:elephant_stationary} shows the probability of the animal being in the exploratory state for January, March, May, July, September and November. Pointwise confidence bands are based on the approximate Bayesian posterior of $\Hat{\bm{\theta}}$ \citep{wood2017generalized} --- corresponding to conditional uncertainty quantification based on the estimated smoothness parameters and the implied effective number of parameters \citep{gray1992flexible}. To get an intuition for the different movement patterns arising from the two states, see Figure \ref{fig:elephant_marginal} in the Appendix.

Our results reveal a clear pattern: the studied elephant is most active during the night and early morning, while remaining largely inactive throughout the day. While the seasonal variation is not extreme --- likely due to the Ivory Coast's proximity to the equator --- it is nonetheless potentially interesting. 
During the hotter summer months, the morning activity peak diminishes and the elephant appears to become active slightly later in the afternoon. These findings highlight the importance of accounting for seasonal effects, while such effects may be much stronger when studying species that inhabit regions further from the equator.

\subsection{Drophila melanogaster --- function-valued random effects}

To demonstrate how function-valued random effects can be incorporated into HMMs by representing them as a tensor-product interaction, we investigate the diel variation of locomotor activity in common fruitflies (\textit{Drosophila melanogaster}) which have been known to synchronise their circadian clocks to the common light-dark cycles for fitness \citep{beaver2002loss, bernhardt2020life}.

The data were collected by \citet{coculla2025hsp90} to investigate individual heterogeneity in the flies' circadian clocks. 
1- to 5-day old male flies were trained under a standard 12-hour-light and 12-hour-dark condition (LD) for 3 days and subsequently exposed to 5 days of consequtive darkness (DD).
To measure the flies' activity, their movement was tracked by counting how often a fly passed an infrared beam in the middle of the tube they were kept in. These counts were aggregated over 30-minute windows, leading to 48 observed counts per fly per day, ranging from 0-335.

For this case study, we compare the wild type to one of the 13 originally studied modified genotypes, namely 
$\text{Hsp83}^{\text{e6A}}$ / $\text{Hsp83}^{\text{08445}}$. 
The two groups contain 35 and 34 individuals, respectively. In total, the data comprises 13,104 observations of the wild-type and 13,056 observations of the modified geneotype. 

The aim is to quantify inter-individual differences in behavioural diel variation. 
We model the data using a 2-state HMM, where the hidden states represent inactive and active behaviour, to account for the noisy measurement of activity through the count observations.
The state-dependent distributions are chosen as negative-binomial distributions to account for potential overdispersion in the activity counts, similar to \citet{feldmann2023flexible, coculla2025hsp90}.
To quantify the inter-individual variation, the transition probabilities are expressed as a function of the time of day, including a main effect as well as an individual-specific function-valued random effect for each of the two light schedules.
Hence, the transition probabilities for animal $a$ 
and condition $c \in \{\text{\textit{LD}}, \text{\textit{DD}}\}$ are modelled as
$$
\text{logit}(\gamma_{a, c, t}^{(ij)}) = \beta_{0,c}^{(ij)} + \alpha_{a,c}^{(ij)} + f_c^{(ij)}(\text{tday}_t) + f_{c,a}^{(ij)}(\text{tday}_t), \quad i \neq j.
$$
By expressing the function-valued random effect using the ANOVA decomposition of a tensor-product interaction, separate variance parameters need to be estimated for the random intercepts $\alpha_{a,c}^{(ij)}$ and the ``random-effect dimension'' of $f_{c,a}^{(ij)}(\text{tday}_t)$. Indeed, this is desirable for biological understanding: The first variance parameter quantifies the variation in the overall activity level, also called \textit{behavioural type}, while the second variance parameter quantifies heterogeneity in the behavioural response to the time of day, also called \textit{behavioural plasticity} \citep{hertel2020guide}.

\begin{figure}
    \centering
    \includegraphics[width=1\linewidth]{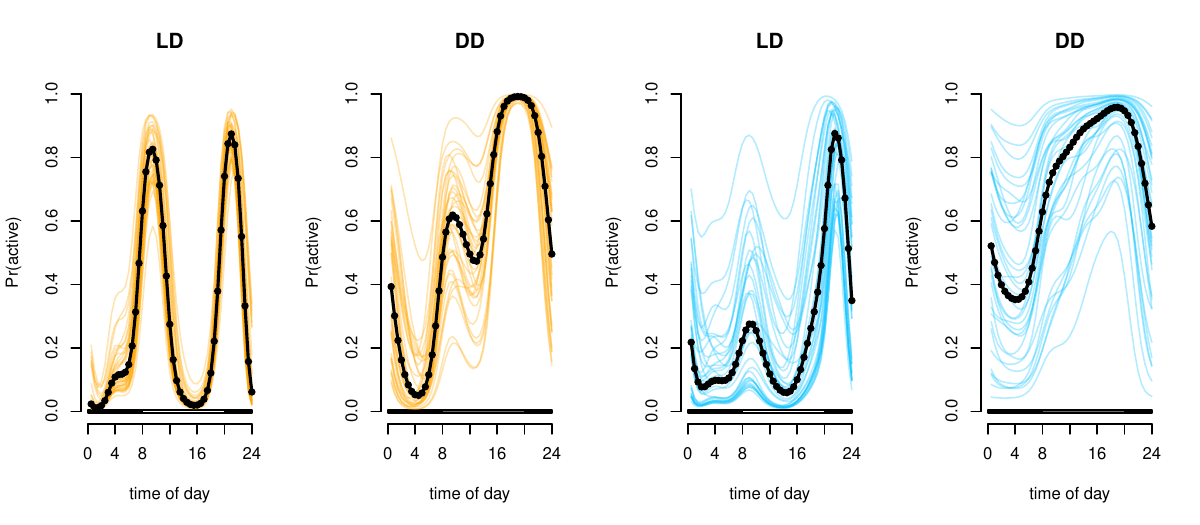}
    \caption{Probability of being active (based on the periodically stationary distribution) as a function of the time of day for both light schedules for the fruit flies of the wild-type (panels 1 and 2) and the modified genotype (panels 3 and 4). The thin coloured lines correspond to the predicted individual-specific effects while the thick dotted line is the main effect.}
    \label{fig:fruitflies_stationary}
\end{figure}

The model specification above results in 16 smoothing parameters that need to be estimated --- four for each tensor-product interaction, and one such interaction for each condition and each off-diagonal entry of the t.p.m. However, for parsimony and better interpretability, we use the mapping functionality detailed in Section \ref{subsec:practical} to estimate the same smoothing parameters for both off-diagonal entries of the t.p.m., corresponding to transitions from inactive to active and vice versa, 
thereby also facilitating simpler group comparisons.

Again, the relevant model matrices were obtained using \texttt{mgcv} with the formula
\begin{center}
    \texttt{condition + \\
s(aniID, bs="re", by=condition) + s(tod, bs="cc", by=condition, k=10) +\\
ti(aniID, tod, bs=c("re","cc"), by=condition, k=c(nAnimals,10))}
\end{center}
We estimate two separate models, one for each genotype, using the extended Fellner-Schall method. 
The models comprise 1268 and 1232 coefficients to be estimated within each penalised fit. Model estimation took 2.5 and 2.3 hours on an Apple M2 chip with 16 GB of memory, requiring a total of 32 and 29 outer iterations for convergence.

The fitted models distinguish between low and high activity with state-dependent mean counts of 0.8 and 44.8 for the wild-type model and 0.3 and 21.7 for the modified-genotype model (see Figure \ref{fig:fruitflies_marginal} in the Appendix).
Figure \ref{fig:fruitflies_stationary} shows each fly's probability of being active based on the predicted periodically stationary distribution \citep{koslik2023inference} for the wild-type and modified genotype and both light conditions, as well as the main effect. In light of this figure, it is evident that the estimated random effects are not purely additive, but the difference betweeen individuals varies with the time of day.

\begin{table}
    \centering
    \begin{tabular}{lcccc}
    \toprule
        & \multicolumn{2}{c}{\textit{LD}} & \multicolumn{2}{c}{\textit{DD}} \\
        & wild-type & modified & wild-type & modified \\
    \midrule
        behavioural type & 0.33 (0.06) & 0.96 (0.17) & 0.57 (0.10) & 1.13 (0.19) \\
        behavioural plasticity & 1.55 (0.14) & 1.62 (0.15) & 1.69 (0.16) & 1.58 (0.15) \\
    \bottomrule
    \end{tabular}
    \caption{Estimated variances for the wild-type and modified genotype under both light schedules. Approximate standard deviations obtained as described in Section \ref{subsec:practical} in brackets.}
    \label{tab:re_variacens_flies}
\end{table}

Additionally, Table \ref{tab:re_variacens_flies} shows the estimated variance parameters (inverse penalty strengths) for the random intercept and functional random effect for both genotypes and light schedules. 
The modified genotype admits a larger variation in the random intercept, while the variances of the function-valued random effects are very similar across the two groups.

\subsection{Arctic muskox --- space-time interaction}

In this last case study, we examine the movement of Arctic muskoxen (\textit{Ovibos moschatus}) to explore potential space-time interactions in the animals' behavioural patterns. The data were collected by \citet{beumer2020application} in 2013 and 2015 and comprise tracks of 19 adult female muskoxen. As the muskoxen's behaviour substantially differs between moving on snow-covered or snow-free ground, the data were split into snow-cover or snow-free. 
For simplicity, here we only analyse the snow-cover data, yielding a total number of 214,213 hourly GPS locations. 
The preprocessed data set contains hourly step lengths and turning angles, which we use for the subsequent analysis.
Building on the results of \citet{beumer2020application} and other previous analyses by \citet{pohle2022flexible} and \citet{koslik2025hidden}, we also use a 3-state model where the states are interpreted as proxies for \textit{resting}, \textit{foraging} and \textit{travelling/relocating} behaviour. The step lengths and turning angles are modelled using state-dependent gamma and von Mises distributions, respectively.

The main question we aim to address is whether the animals' locations influence their behavioural decisions. Specifically, do certain locations imply a higher probability for the animals to initiate the foraging state?
This would indicate spatial preferences for foraging. A subsequent question is then if such a spatial effect is constant or does indeed vary temporally. 
Thus, to address this, we express the transition probability from travelling to foraging as a smooth function of the position and time of day, because this transition corresponds to the end of relocation due to a suitable foraging patch being found. Specifically, we again choose the ANOVA decomposition

$$\gamma_{3,2}^{(t)} = \beta_0^{(3,2)} + f(x_t, y_t) + f(\text{tday}_t) + f(x_t, y_t, \text{tday}_t).$$

Additionally, we fit a homogeneous model, models comprising only a spatial effect \textit{or} a temporal effect, and a model comprising an additive spatiotemporal effect.

Similar to the other case studies, all model matrices were obtained using \texttt{mgcv} with the formula
\begin{center}
\texttt{s(x, y, bs = "tp", k = 50) + s(tday, bs = "cc", k = 8) + \\
ti(x, y, tday, d = c(2,1), bs = c("tp", "cc"), k = c(50, 8))}
\end{center}

The spatiotemporal model comprised 364 total coefficients and estimation of four smoothness parameters was necessary. Model fitting took 9.2 hours on an Apple M2 chip with 16 GB of memory, requiring a total of 19 outer iterations for convergence. 

\begin{table}[]
    \centering
    \begin{tabular}{lccc}
    \toprule
        transition probability formula & log-likelihood & $\Delta$ AIC & $\Delta$ BIC\\
    \midrule
        $\sim$ \texttt{1} & -1385002 & 495.45 & 66.85\\
        $\sim$ \texttt{s(x,y)} & -1384798 & 165.05 & 133.54\\
        $\sim$ \texttt{s(tday)} & -1384932  & 368.04 & \textbf{0.00}\\
        $\sim$ \texttt{s(x,y) + s(tday)} & -1384711 & 2.91 & 35.02\\
        $\sim$ \texttt{s(x,y) + s(tday) + s(x,y,tday)} & -1384698 & \textbf{0.00} & 147.57 \\
    \bottomrule
    \end{tabular}
    \caption{Log-likelihood values and information criteria of the five models fitted to the muskox data.}
    \label{tab:muskox_IC}
\end{table}

Information criteria for all candidate models are reported in Table \ref{tab:muskox_IC}. 
We find that AIC very slightly prefers the full interaction model over the simpler, additive model of spatial and temporal variation, while BIC prefers the much simpler model, only comprising temporal variation.
The estimated state-dependent distributions (based on the homogeneous model) confirm the interpretation of the states as resting, foraging, and travelling (see Figure \ref{fig:muskox_marginal} in the Appendix).

Figure \ref{fig:muskox_spatial} shows the estimated spatial effect on the probability of transitioning from the travelling mode to foraging at 9 AM and 6 PM, obtained from the spatiotemporal model. While the temporal variation is not very pronounced, some slight differences between the two time points can be observed, and some areas seem to be preferred for initiating foraging behaviour.
It is highly likely that the fitted smooth function effectively serves as a proxy for unexplained variation induced by unmeasured covariates such as habitat quality. The fitted smooth function then does not reveal the actual source of the spatially varying state-switching dynamics, nevertheless the spatial (or spatiotemporal) smooth can still improve the model fit and hence be valuable for predictive purposes. For example, such a spatial model can help to more precisely identify foraging patches, resting areas or breeding zones, in the spatiotemporal model additionally including year-round variation of these activities at given sites, all of which can aid conservation efforts.


\begin{figure}
    \centering
    \includegraphics[width=1\linewidth]{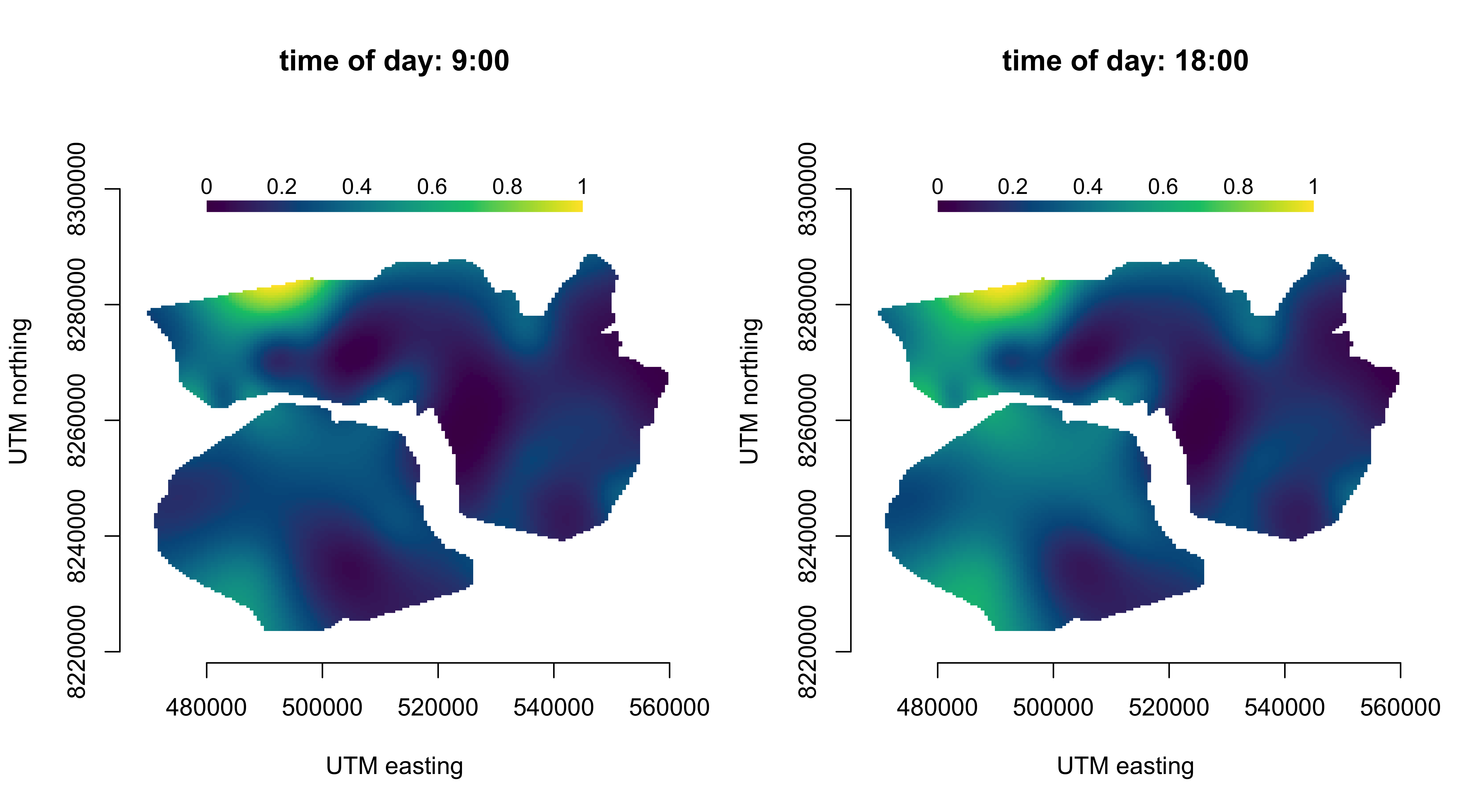}
    \caption{Estimated probability of the muskox transitioning from the travelling to the foraging state as a function of the location in the study region at 9 AM and 5 PM. Cells outside the study region are left blank.}
    \label{fig:muskox_spatial}
\end{figure}


\section{Discussion and outlook}


We have developed a robust and computationally efficient approach for automatic smoothness selection in nonparametric Markov-switching models that incorporate tensor-product interactions. By combining the extended Fellner–Schall method with automatic differentiation and fast recursive likelihood evaluation, we enable practical estimation of models that were previously computationally prohibitive. 

Our implementation leverages the flexibility of the \texttt{RTMB} package to provide user-friendly software tools that seamlessly integrate arbitrary tensor-product interactions into custom Markov-switching models, supporting user-defined likelihood functions written in simple \texttt{R} code. This modularity makes our approach widely accessible for applied researchers.
While this study focused on discrete-time Markov-switching models, the procedure is broadly applicable to a wider range of latent Markovian models, including continuous-time and continuous-space models \citep{mews2024build} as well as hidden semi-Markov models \citep{langrock2011hidden, koslik2025hidden}. 

Through three case studies, we demonstrated the flexibility of tensor-product interactions to capture complex covariate effects such as bivariate smoothing, function-valued random effects, and space–time interactions. 
While the case studies have shown that the increased model complexity entailed by these smooths might not always be necessary, it is nevertheless valuable for practitioners to be able to fit such much more detailed models whenever appropriate. 
Combined with the ever-increasing availability of high-resolution sensor data, this framework opens new avenues to uncover subtle behavioural dynamics, individual heterogeneity, and environmental influences that may have been obscured by simpler models.
This empowers researchers to apply a large variety of modern smoothing techniques and paves the way for more nuanced inference in complex ecological systems and beyond.




\section*{Supplementary materials}
The data and code for fully reproducing all case studies can be found at \url{https://github.com/janoleko/tp_interactions}. The code for the extended Fellner-Schall procedure can be found at \url{https://github.com/janoleko/LaMa/blob/main/R/qreml_functions.R}

\section*{Acknowledgments}

The author is very grateful to Angelica Coculla and Ralf Stanewsky for providing the Drosophila melanogaster activity data and sincerely thanks Roland Langrock, Thomas Kneib, and Carlina Feldmann for their helpful comments on an earlier version of this manuscript.

\bibliographystyle{agsm}  
\bibliography{references}

\newpage

\appendix
\renewcommand{\thesubsection}{\Alph{section}.\arabic{subsection}}

\section{Appendix}

\begin{figure}[h]
    \centering
    \includegraphics[width=1\linewidth]{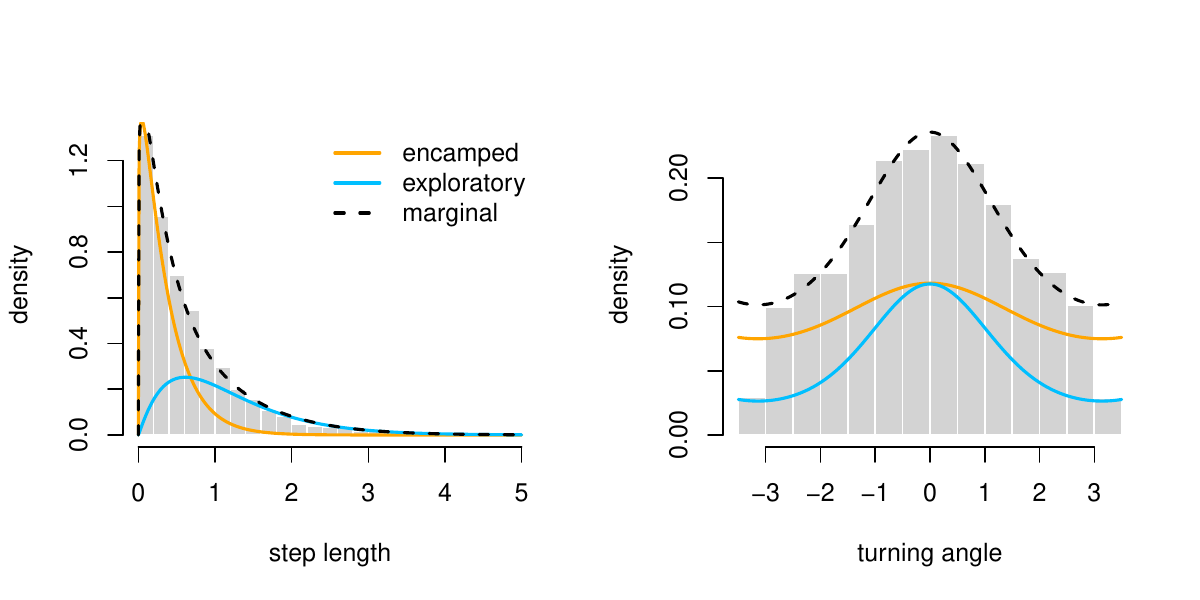}
    \caption{Weighted state-dependent step-length (left panel) and turning angle (right panel) distributions in the encamped (orange) and exploratory (light-blue) state, complemented with the marginal distribution (black).}
    \label{fig:elephant_marginal}
\end{figure}

\begin{figure}[h]
    \centering
    \includegraphics[width=1\linewidth]{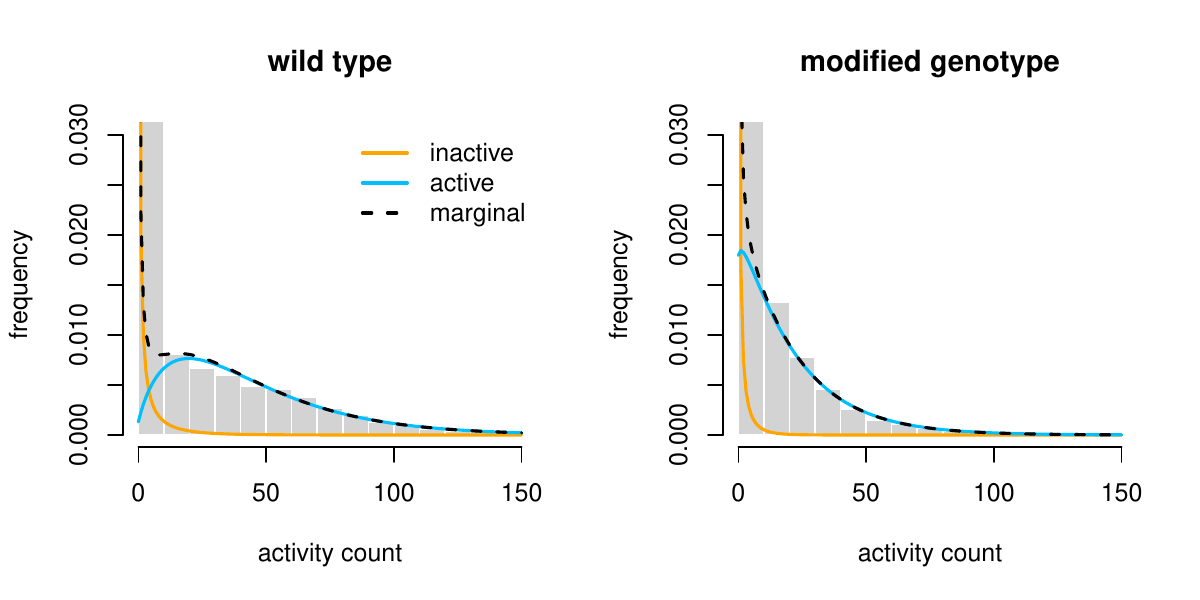}
    \caption{Weighted state-dependent negative binomial distributions for the wild type (left panel) and the modified genotype (right panel) in the inactive (orange) and active (light-blue) state, complemented with the marginal distribution (black). The discrete probability mass functions are displayed like densities for visual clarity.}
    \label{fig:fruitflies_marginal}
\end{figure}

\begin{figure}[h]
    \centering
    \includegraphics[width=1\linewidth]{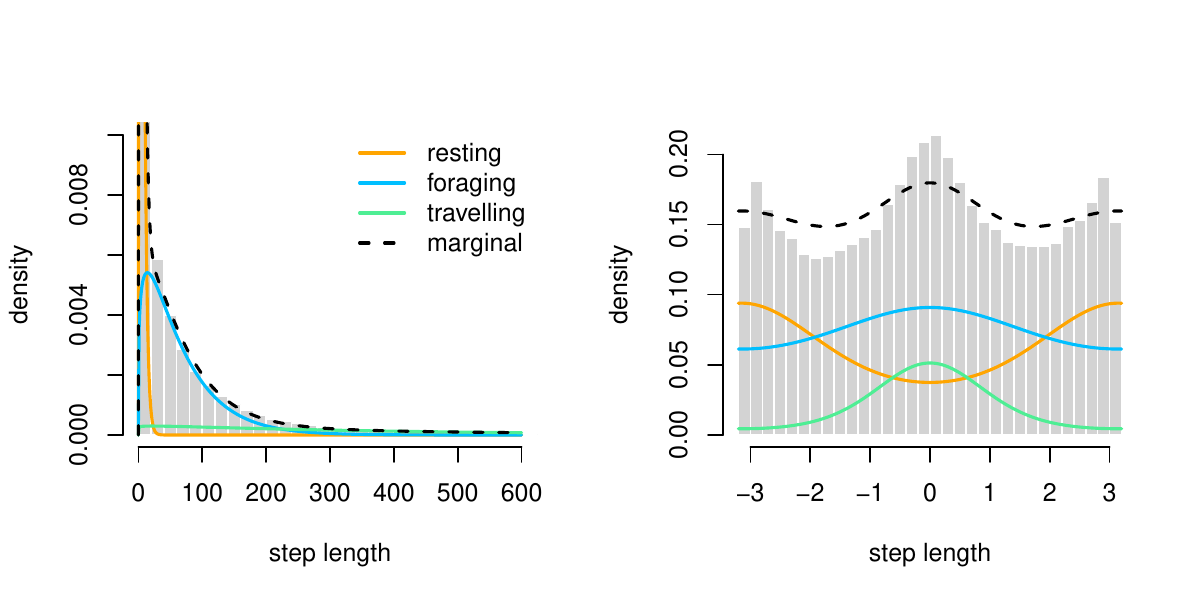}
    \caption{Weighted state-dependent step-length (left panel) and turning angle (right panel) distributions in the resting (orange) and foraging (light-blue), and travelling (green) state, complemented with the marginal distribution (black).}
    \label{fig:muskox_marginal}
\end{figure}

\end{document}